# Virtualization Implementation Model for Cost Effective & Efficient Data Centers

Mueen Uddin[1]

Department of Information Systems,
Universiti Teknologi Malaysia
Mueenmalik9516@gmail.com

Azizah Abdul Rahman[2]

Department of Information Systems,
Universiti Teknologi Malaysia
azizahar@utm.my

**ABSTRACT: -** Data centers form a key part of the infrastructure upon which a variety of information technology services are built. They provide the capabilities of centralized repository for storage, management, networking and dissemination of data. With the rapid increase in the capacity and size of data centers, there is a continuous increase in the demand for energy consumption. These data centers not only consume a tremendous amount of energy but are riddled with IT inefficiencies. Data center are plagued with thousands of servers as major components. These servers consume huge energy without performing useful work. In an average server environment, 30% of the servers are "dead" only consuming energy, without being properly utilized. This paper proposes a five step model using an emerging technology called virtualization to achieve energy efficient data centers. The proposed model helps Data Center managers to properly implement virtualization technology in their data centers to make them green and energy efficient so as to ensure that IT infrastructure contributes as little as possible to the emission of greenhouse gases, and helps to regain power and cooling capacity, recapture resilience and dramatically reducing energy costs and total cost of ownership.

## I.    INTRODUCTION

Data centers form the backbone of a wide variety of services offered via the Internet including Web-hosting, e-commerce, social networking, and a variety of more general services such as software as a service (SAAS), platform as a service (PAAS), and grid/cloud computing [1]. They consist of concentrated equipment to perform different functions like Store, manage, process, and exchange digital data and information. They support the informational needs of large institutions, such as corporations and educational institutions, and provide application services or management for various types of data processing, such as web hosting, Internet, intranet, telecommunication, and information technology [2]. Data centers are found in nearly every sector of the economy, ranging from financial services, media, high-tech, universities, government institutions, and many others. They use and operate data centers to aid business processes, information management and communication functions [3]. Due to rapid growth in the size of the data centers there is a continuous increase in the demand for both the physical infrastructure and IT equipment, resulting in continuous increase in energy consumption.

Data center IT equipment consists of many individual devices like Storage devices, Servers, chillers, generators, cooling towers and many more. Servers are the main consumers of energy because they are in huge number and their size continuously increases with the increase in the size of data centers. This increased consumption of energy causes an increase in the production of greenhouse gases which are hazardous for environmental health.

Virtualization technology is now becoming an important advancement in IT especially for business organizations and has become a top to bottom overhaul of the computing industry. Virtualization combines or divides the computing resources of a server based environment to provide different operating environments using different methodologies and techniques like hardware and software partitioning or aggregation, partial or complete machine simulation, emulation and time sharing [4].

It enables running two or more operating systems simultaneously on a single machine. Virtual machine monitor (VMM) or hypervisor is a software that provides platform to host multiple operating Systems running concurrently and sharing different resources among each other to provide services to the end users depending on the service levels defined before the processes. Virtualization and server consolidation techniques are proposed to increase the utilization of underutilized servers so as to decrease the energy consumption by data centers and hence reducing the carbon footprints.

This paper identifies some of the necessary requirements to be fulfilled before implementing virtualization in any firm. Section 2 describes a complete description about data centers. Section 3 emphasizes the need for implementing virtualization technology in a data center and provides a five step model of implementing it. Section 4 discusses some of the advantages of virtualization after being implemented. In the end conclusions and recommendations are given.

## II. PROBLEM STATEMENT

Data Centers are the main culprits of consuming huge energy and emitting huge amount of CO2, which is very hazardous for global warming. Virtualization technology provides the solution but it has many overheads, like single point of failure, total cost of ownership, energy and efficiency calculations and return of investment. The other problem faced by IT managers related with the proper implementation of Virtualization technology in data centers to cop up with the above defined problems. This paper comes up with a model to be followed by IT managers to properly implement virtualization in their data centers to achieve efficiency and reduce carbon footprints.

## III. LITERATURE REVIEW

In recent years the commercial, organizational and political landscape has changed fundamentally for data center operators due to a confluence of apparently incompatible demands and constraints. The energy and environmental impact of data centers has recently become a significant issue for both operators and policy makers. Global warming forecasts that rising temperatures, melting ice and population dislocations due to the accumulation of greenhouse gases in our atmosphere from use of carbon-based energy. Unfortunately, data centers represent a relatively easy target due to the very high density of energy consumption and ease of measurement in comparison to other, possibly more significant areas of IT energy use. Policy makers have identified IT and specifically data centre energy use as one of the fastest rising sectors. At the same time the commodity price of energy has risen faster than many expectations. This rapid rise in energy cost has substantially impacted the business models for many data centers. Energy security and availability is also becoming an issue for data centre operators as the combined pressures of fossil fuel availability, generation and distribution infrastructure capacity and environmental energy policy make prediction of energy availability and cost difficult [5].

As corporations look to become more energy efficient, they are examining their operations more closely. Data centers are found a major culprit in consuming a lot of energy in their overall operations. In order to handle the sheer magnitude of today's data, data centers have grown themselves significantly by continuous addition of thousands of servers. These servers are consuming much more power, and have become larger, denser, hotter, and significantly more costly to operate [6]. An EPA Report to Congress on Server and Data Center Energy Efficiency completed in 2007 estimates that data centers in USA consume 1.5 percent of the total USA electricity consumption for a cost of $4.5 billion [7]. From the year 2000 to 2006, data center electricity consumption has doubled in the USA and is currently on a pace to double again by 2011 to more than 100 billion kWh, equal to $7.4 billion in annual electricity costs [8].

Gartner group emphasizes on the rising cost of energy by pointing out that, there is a continuous increase in IT budget from 10% to over 50% in the next few years. Energy increase will be doubled in next two years in data centers [9]. The statistics Cleary shows that the yearly cost of power and cooling bill for servers in data centers are around $14billion and if this trend persists, it will rise to $50billion by the end of decade [10].

With the increase in infrastructure and IT equipment, there is a considerable increase in the energy consumption by the data centers, and this energy consumption is doubling after every five years. [11]. Today's data centers are big consumer of energy and are filled with high density, power hungry equipment. If data center managers remain unaware of these energy problems then the energy costs will be doubled between 2005 and 2011. If these costs continue to double every five years, then data center energy costs will increase to 1600 % between 2005 and 2025 [12]. Currently USA and Europe have largest data center power usage but Asia pacific region is rapidly catching up. [13].

## IV. PROPOSED WORK

This paper comes up with a virtualization implementation model to be followed by IT managers before implementation of Virtualization technology in their data centers. This technology has many overheads like single point of failure, total cost of ownership, energy and efficiency calculations and return of investment. In this paper I proposed a five step model can be called as pre requisites for the proper implementation of virtualization. The proposed model signifies the importance of categorizing the resources of data center into different resource pools and then selecting and applying the proper virtualization where needed, to maximize the performance of different components as whole in the data center.

Before implementing server virtualization in any firm it is important to seriously plan and consider virtualization risks associated with it. It is also important for the data center to check whether it has the necessary infrastructure to handle the increased power and cooling densities arise due to the implementation of virtualization. It is also important to consider the failure of single consolidated server, because it is handling the workload of multiple applications. In order to properly implement virtualization there is a need to answer some of the questions:

- What is virtualization?
- Why we need it?
- How it can improve our businesses?
- Types of virtualization technologies exist?
- What is cost/benefit ratio of virtualization?
- What new challenges it will bring to business firms?
- Structure of virtualization solution being implemented?
- Which applications or services are good virtualization candidates?
- Server platforms best suited to support virtualization?

### A. Proposed Model

We define a layered model consisting of five layers and further each layer comprising of more detailed processes. These components provide a detailed treatment of state of the art and emerging challenges faced by data centers managers to implement and manage virtualization properly in their data centers to achieve desired objectives. The proposed model defines that, the process of virtualization should be structured and designed in such a way that it must fulfill the necessary requirements and should be within the scope & infrastructure domain already installed in the data center. It is therefore much more than simply loading a virtualization technology on different servers and transforming one or two workloads into virtual machines. Rather it is a complex and rigorous process that need to be implemented and monitored properly. The proposed model defines five key steps need to be followed at different stages in a structural way to achieve the efficiency required in the data center. The components of proposed model are listed below:

- Inventory Process
- Type & Nature of Virtualization
- Hardware Maximization
- Architecture
- Manage Virtualization

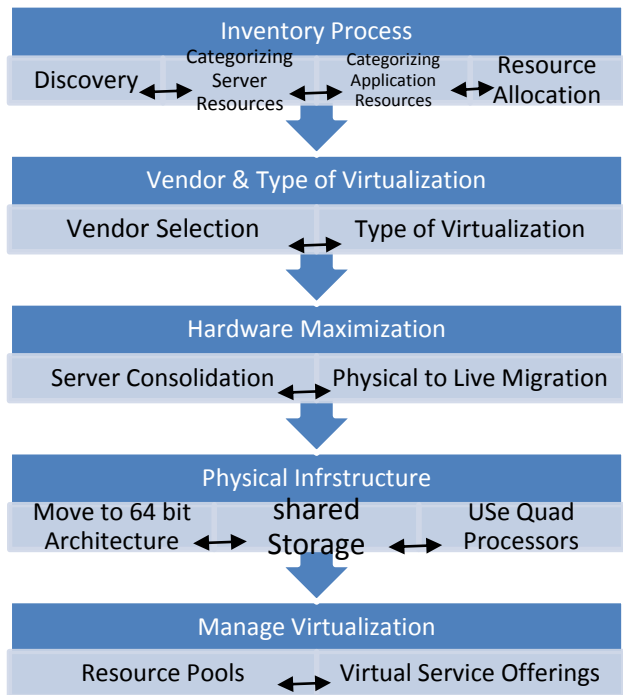

Figure1. Process of Virtualization

### B. Inventory Process

The process of virtualization starts by creating an inventory of all hardware and software resources including servers and their associated resources and workloads they require for processing, storage components, networking components etc. The inventory process includes both utilized and idle servers. This process also includes information related to:

- Make and Model of the Processor
- Types of processors (socket, Core, Threads, Cache)
- Memory size and speed
- Network type (Number of ports, speed of each port)
- Local storage (number of disk drives, capacity, RAID)
- Operating system and their patch levels (service levels)
- Applications installed
- Running services
- Running Application
- Storage Devices etc.

The inventory process also discovers, identifies and analyzes an organizations network before it is being virtualized. It consists of following phases:

a) Discovery: It is very important for an organization to know in advance the total content of its infrastructure before implementing virtualization. This is the most important step in any virtualization project. There are many tools available from different vendors for performing initial analysis of an organization.

Microsoft Baseline Security Analyzer (MBSA) tool provides different information like IP addressing, Operating System, installed applications and most importantly vulnerabilities of every scanned system. After analyzing, all generated values are linked to MS Visio, which generates a complete inventory diagram of all components and also provides details about each component being analyzed. Microsoft Assessment and Planning toolkit (MAP) is another tool for the assessment of network resources. It works with windows management instrumentation (WMI), the remote registry service or with simple network management protocol to identify systems on network. VMware, the founder of X-86 virtualization, also offers different tools for the assessment of servers that could be transformed into virtual machines. VMware Guided Consolidation (VGC) a powerful tool assesses network with fewer than 100 physical servers. Since VGC is an agent less tool it doesn't add any overhead over production server's workload.

b) Categorize Server Resources: After creating server inventory information, the next step is to categorize the servers and their associated resources and workloads into resource pools. This process is performed to avoid any technical political, security, privacy and regulatory concern between servers, which prevent them from sharing resources. Once analysis is performed, we can categorize each server roles into groups. Server roles are categorized into following service types:

- Network infrastructure servers
- Identity Management servers
- Terminal servers
- File and print servers
- Application servers
- Dedicated web servers
- Collaboration servers
- Web servers
- Database servers

c) Categorizing Application Resources: After categorizing servers into different resource pools, applications will also be categorized as:

- Commercial versus in-house
- Custom applications
- Legacy versus updated applications
- Infrastructure applications
- Support to business applications
- Line of business applications
- Mission critical applications

d) Allocation of Resources: After creating the workloads, the next process is to allocate computing resources required by these different workloads and then arranging them in normalized form, but for normalization the processor utilization should be at least 50%. It is very important to normalize workloads so as to achieve maximum efficiency in terms of energy, cost and utilization. The formula proposed in this paper for normalization is to multiply utilization ratio of each server by total processor capacity that is (maximum processor efficiency * number of processors * number of cores).

### C. Type & Nature of Virtualization

After analyzing and categorizing servers and other resources, the second step defines virtualization in more detail like its advantages, its types, layers and most importantly vendor identification and selection whose product most suits and fulfills all criteria for data gathered in first step.

VMware Capacity Planner (VCP) tool can be used when network size extends over 100 physical servers. It generates reports on server processor utilization including CPU, Memory, and network and disk utilization on server by server basis and finally identifies potential virtualization candidates. Other tools like CIRBA's Power Recon and Plate Spin's are also very useful tools which analyze technical and non-technical factors in data centers and generate reports for the consolidation of servers. It should be noted that all analysis should be done on time for a period of at least one month; this will generate high and low utilization ratios for each server.

### D. Hardware Maximization

This is the most important step of virtualization process. Since servers are now going to run multiple virtual workloads, it is important to consider hardware issues because already available hardware is not enough and suitable for providing high availability of virtual workloads. A change is required to install new hardware that supports and delivers the best price and performance. This process ensures high availability of virtual workloads and also provides leaner and meaner resource pool of resources for these virtual workloads. Hardware maximization can also be achieved by purchasing new quad core processors which have better hardware utilization capability along with less consumption of energy hence emissions of $CO_2$ is greatly reduced.

a) Serer Consolidation: Server consolidation is an approach to efficiently use Server resources in order to reduce the total number of servers or server locations to maximize the hardware utilization. This technique is used to overcome the problems of server sprawl, a situation in which multiple, under-utilized servers take up more space and consume more resources than can be justified by their workload. The process of server consolidation always begins from servers that are mostly underutilized and remain idle for long durations of time. The other most important reason for applying server consolidation is that these servers are in huge quantity while the available resources are very much limited. Servers in many companies typically run at 15-20% of their capacity, which may not be a sustainable ratio in the current economic environment. Businesses are increasingly turning to server consolidation as one means of cutting unnecessary costs and maximizing return on investment (ROI) in the data centers [5].

b) Physical to Virtual Live Migration: Physical to Live Migration is the most critical, time-consuming and painful operation when performed manually. Mostly data center managers find this process much more complex and rigorous. It includes cloning existing operating system and restoring it on an identical machine, but at the same time changing the whole underlying hardware, which can lead to driver reinstallation or possibly the dreadful blue screen of death. To avoid these ambiguities, virtualization vendors started to offer different physical to virtual (P2V) migration utilities. These software's speeds up the movement of operation and solve on the fly driver incompatibilities, by removing physical hardware dependencies from server operating systems and allowing them to be moved and recovered. Instead of having to perform scheduled hardware maintenance at some obscure hour over the weekend, server administrators can now live migrate a VM to another physical resource and perform physical server hardware maintenance in the middle of the business day. Virtuozzo for Windows 3.5.1 SWsoft itself introduced a physical to virtual (P2V) migration tool called VZP2V. This tool can remotely install P2V knowing machine administrative username and password.

### E. Physical Infrastructure

Data Center physical infrastructure is the foundation upon which Information Technology and telecommunication network resides. It is the backbone of businesses, and its elements provide the power, cooling, physical housing, security, fire protection and cablings which allow information technology to function. This physical infrastructure as whole helps to design, deploy and integrate a complete system that helps to achieve desirable objectives [15].

a) Move To 64-Bit Architecture: The architecture of a machine consists of set of different instructions that allow inspecting or modifying machine state trapped when executed in any or most probably the privileged mode. To support proper hardware utilization, it is important to update and revise whole datacenter architecture. To protect virtual workloads, x-64 systems should be linked to shared storage and arranged into some form of high availability clusters so as to minimize the single point of failure. One of the major issues in hardware maximization is the proper utilization and availability of RAM for each virtual machine. For this reason it is important to consider 64 bit architecture, which provides more utilization and availability of RAM for all virtual and physical systems.

b) Rely On Shared Storage: It is also important to consider single point of failure because one server is now running the workloads of multiple servers. If this server goes down the whole process of virtualization becomes fail. To remove the chances of single point of failure at any stage can be achieved by using redundancy and clustering services to protect virtual workloads. These services are mostly provided by Microsoft and Citrix. While VMware on the other hand uses custom configuration approach called High availability (HA).

### F. Manage Virtualization

This is the most important step that involves end users and the top management to make the decision whether to implement the virtualization or not. It involves many factors like cost, return on investment, security, and service level agreements. Virtualized data centers are managed by dividing the functionalities of data center into two layers.

- Resource Pool (RP)
- Virtual Service Offering (VSO)

It is very much important to consider the available resources and what other resources are required to fulfill the requirements of proper implementation of virtualization in data center. It is also important to note that conversion should always be preferred when servers are offline to protect existing services and maintain service level agreements (SLA) with end users.

### V. BENEFITS OF VIRTUALIZATION

Virtualization promises to radically transform computing for the better utilization of resources available in the data center reducing overall costs and increasing agility. It reduces operational complexity, maintains flexibility in selecting software and hardware platforms and product vendors. It also increases agility in managing heterogeneous virtual environments. Some of the benefits of virtualization are

### A. Server & Application Consolidation

Virtual machines can be used to consolidate the workloads of under-utilized servers on to fewer machines, perhaps a single machine. The benefits include savings on hardware and software, environmental costs, management, and administration of the server infrastructure. The execution of legacy applications is well served by virtual machines. A legacy application may not run on newer hardware or operating systems. Even if it does, it may under-utilize the server, hence virtualization consolidates several such applications, which are usually not written to co-exist within a single execution environment. Virtual machines provide secure, isolated sandboxes for running entrusted applications.

Examples include address obfuscation. Hence Virtualization is an important concept in building secure computing platforms.

### B. Multiple Execution Environments

Virtual machines can be used to create operating systems or execution environments that guarantee resource management by using resource management schedulers with resource limitations. Virtual machines provide the illusion of hardware configuration such as SCSI devices. It can also be used to simulate networks of independent computers. It enables to run multiple operating systems simultaneously having different versions, or even different vendors.

### C. Debugging and Performance

Virtual machines allow powerful debugging and performance monitoring tools that can be installed in the virtual machine monitor to debug operating systems without losing productivity. Virtual machines provide fault and error containment by isolating applications and services they run. They also provide behavior of these different faults. Virtual machines aid application and system mobility by making software's easier to migrate, thus large application suites can be treated as appliances by "packaging" and running each in a virtual machine. Virtual machines are great tools for research and academic experiments. They provide isolation, and encapsulate the entire state of a running system. Since we can save the state, examine, modify and reload it. Hence it provides an abstraction of the workload being run.

### D. Resource Sharing

Virtualization enables the existing operating systems to run on shared memory multiprocessors. Virtual machines can be used to create arbitrary test scenarios, and thus lead to very imaginative and effective quality assurance. Virtualization can also be used to retrofit new features in existing operating systems without "too much" work. Virtualization makes tasks such as system migration, backup, and recovery easier and more manageable. Virtualization provides an effective means of binary compatibility across all hardware and software platforms to enhance manageability among different components of virtualization process.

### VI. CONCLUSION

This paper highlights the importance of virtualization technology being implemented in data centers to save the cost and maximize the efficiency of different resources available. We proposed a five step model to properly implement virtualization. It starts by categorizing servers and their associated applications and resources into different resource pools. It is important to consider that virtualization not only needs to characterize the workloads that are planned to be virtualized, but also target the environments into which the workloads are to be applied. It is important to determine the type of servers, their current status whether idle or busy, how much it will cost to implement server virtualization, the type of technology needed to achieve the service levels required and finally meet the security/privacy objectives. It is also important for the data center to check whether it has the necessary infrastructure to handle the increased power and cooling densities arise due to the implementation of virtualization.

It is also important to consider the failure of single consolidated server, because it is handling the workload of multiple applications. It poses many challenges to the data center physical infrastructure like dynamic high density, under-loading of power/cooling systems, and the need for real-time rack-level management. These challenges can be met by row-based cooling, scalable power and predictive management tools. These solutions are based on design principles that simultaneously resolve functional challenges and increase efficiency.

## AUTHOR BIOGRAPHIES

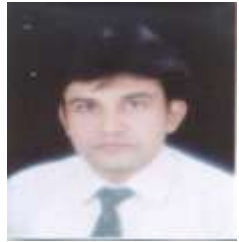

**Mueen Uddin** is a PhD student at University Technology Malaysia UTM. His research interests include digital content protection and deep packet inspection, intrusion detection and prevention systems, analysis of MANET routing protocols, green IT, energy efficient data centers and Virtualization. Mueen has a BS & MS in Computer Science from Isra University Pakistan with specialty in Information networks. Mueen has over ten international publications in various journals and conferences.

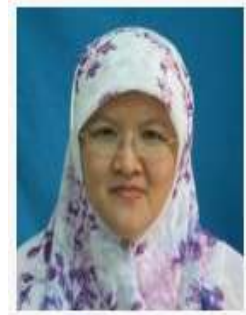

**Azizah Abdul Rahman** is an Associate Professor at University Technology Malaysia. His research interests include designing and implementing techniques for information systems in an organizational perspective, knowledge management, designing networking systems in reconfigurable hardware and software, and implementing security protocols needed for E-businesses. Azizah has BS and MS from University Technology Malaysia. She is a member of the IEEE, AIS, ACM, and JIS. Azizah is a renowned researcher with over 40 publications in journals and conferences.